\documentclass[prd,10pt,aps,superscriptaddress,twocolumn,amsmath,amssymb,nofootinbib,longbibliography]{revtex4-1}

\usepackage{natbib}
\usepackage[colorlinks]{hyperref}
\hypersetup{linkcolor=blue,citecolor=blue,filecolor=blue,urlcolor=blue}

\usepackage{graphicx}
\usepackage{dcolumn}
\usepackage{bm}
\usepackage[tikz]{xcolor}
\usepackage{enumerate}
\usepackage{tikz}

\definecolor{green}{rgb}{0.19,0.64,0.54}
\definecolor{blue}{rgb}{0,0,1}
\definecolor{reddish}{rgb}{0.65, 0.2, 0.2}
\definecolor{darkgreen}{rgb}{0.2,0.7,0.3}
\definecolor{darkblue}{rgb}{0.3,0.40,0.48}
\definecolor{gray}{rgb}{.8,.8,.8}

\def\dd{\text{d}}

\begin{document}
\title{Modelling the braking ``index'' of isolated pulsars}

\author{E. C. A. \surname{Araujo}}
\email{erickycaa@gmail.com}

\affiliation{Institute of Myology, 47 Bd de l'Hôpital,
  75013 Paris, France}

\author{V. A. \surname{De Lorenci}}
\email{delorenci@unifei.edu.br}

\affiliation{Instituto de F\'{\i}sica e Qu\'{\i}mica, Universidade
  Federal de Itajub\'a, \\ Itajub\'a, Minas Gerais 37500-903, Brazil}

\affiliation{{${\cal G}\mathbb{R}\varepsilon\mathbb{C}{\cal
      O}$}---Institut d'Astrophysique de Paris, CNRS \& Sorbonne
  Universit\'e, UMR 7095 98 bis Boulevard Arago, 75014 Paris, France}

\author{P. \surname{Peter}}
\email{peter@iap.fr}
\affiliation{{${\cal G}\mathbb{R}\varepsilon\mathbb{C}{\cal
      O}$}---Institut d'Astrophysique de Paris, CNRS \& Sorbonne
  Universit\'e, UMR 7095 98 bis Boulevard Arago, 75014 Paris, France}

\author{L. S. \surname{Ruiz}}
\email{lucasruiz@unifei.edu.br}
\affiliation{Instituto de Matem\'atica e Computa\c{c}\~ao,
  Universidade Federal de Itajub\'a, \\ Itajub\'a, Minas Gerais
  37500-903, Brazil}

\affiliation{CFisUC, Departamento de Física, Universidade de Coimbra,
  3004-531, Coimbra, Portugal }

\begin{abstract}

An isolated pulsar is a rotating neutron star possessing a high
magnetic dipole moment that generally makes a finite angle with its
rotation axis. As a consequence, the emission of magnetic dipole
radiation (MDR) continuously takes away its rotational energy. This
process leads to a time decreasing angular velocity of the star that
is usually quantified in terms of its braking index. While this simple
mechanism is indeed the main reason for the spin evolution of isolated
pulsars, it may not be the only cause of this effect.  Most of young
isolated pulsars present braking index values that are consistently
lower than that given by the MDR model. Working in the weak field
(Newtonian) limit, we take in the present work a step forward in
describing the evolution of such a system by allowing the star's shape
to wobble around an ellipsoidal configuration as a backreaction effect
produced by the MDR emission. It is assumed that an internal damping
of the oscillations occurs, thus introducing another form of energy
loss in the system, and this phenomenon may be related to the
deviation of the braking index from the pure MDR model
predictions. Numerical calculations suggest that the average braking
index for typical isolated pulsars can be thus simply explained.

\end{abstract}

		
\maketitle

\section{Introduction}
\label{introduction}

The identification of isolated pulsars \cite{hewish1967} with rotating
neutron stars presenting a high surface magnetic field was suggested
long ago \cite{gold1968}, where predictions about their spin evolution
were also anticipated. Shortly thereafter, the pulsar detected in the
Crab nebula was measured \cite{richards1968} to slow down.  It is by
now well understood that the main cause of the spin evolution of
pulsars is the loss of energy due to the emission of magnetic dipole
radiation \cite{pacini1967,pacini1968,gunn1969}.  As this radiation
carries angular momentum, the star angular velocity decreases with
time, leading to a slowing-down behavior that has been detected for
the last 50 years.

When only this source is considered, the rate at which the rotational
energy is radiated away from the pulsar is described by the well known
equation of dipole radiation \cite{shapiro,pacini1968,gunn1969} (in
units where $c=1=\mu_0$, where $\mu_0$ the magnetic vacuum
permeability)
\begin{equation}
\dot E = -\frac{2}{3}\mu^2\Omega^4\sin^2\!\alpha,
\label{eq1}
\end{equation}
where $\Omega = 2\pi \nu$ is the angular velocity of the star, with
$\nu$ its rotation frequency, while $\mu$ is the magnitude of its
magnetic dipole moment $\boldsymbol{\mu}$, making an angle $\alpha$
with its rotation axes, determined by $\boldsymbol{\Omega}$, as
illustrated in Fig.~\ref{ellipsoide}. In Eq.~\eqref{eq1} and in what
follows, a dot over a physical quantity represents its time
derivative. In our system of units, the amplitude $\mu$ of the
magnetic dipole moment is related with that of the magnetic field $B$
through $\mu = \frac12 B R^3$.

The relationship between $\dot E$, $\Omega$, and the moment of inertia
$I$ of the star is given by the torque equation $\dot \Omega =
(I\Omega)^{-1}\dot E$, leading to $\dot \Omega = - \beta \;\Omega^3$,
where we have defined
\begin{equation}
\beta \doteq \frac23 \frac{\mu^2}{I} \sin^2 \alpha.
\label{beta}
\end{equation}
More generally, the different causes behind the slowing-down
phenomenon can be encapsulated by the power law formula
\cite{goldwire1969} $\dot \Omega = - K \Omega^n$, where $n$ is the
so-called braking ``index'' of the pulsar. It is usually defined as if
$K$ were constant, namely
\begin{equation}
n \doteq \frac{\Omega \ddot\Omega}{\dot\Omega^2},
\label{brake}
\end{equation}
and is actually constant only for constant $K$.

In the oversimplified model above with constant magnetic dipole
moment, angle and moment of inertia, i.e. assuming $K_\text{dipole}
=\beta$ constant, one naturally gets $n=3$, while a model based only
on gravitational waves radiation \cite{ostriker1969} leads to $n=5$.

Precise measurements of the braking index of several isolated pulsars
have revealed values that are consistently different than that
predicted on a purely magnetic dipole model with constant $K$ (see for
instance the review in Ref. \cite{lyne2015}).  Consequences of
assuming $K$ as time dependent function~\cite{Blandford1988} has been
examined~\cite{magalhaes2012,lyne2015}, and depending on the way $K$
evolves in time, the braking index can be $n\not=3$.  Furthermore,
assuming~\cite{hamil2015} $I = I[\Omega(t)]$ only in the magnetic
dipole model is not enough to explain the present observational data
for the known isolated pulsar. However, it was
reported~\cite{hamil2016} that assuming a time dependent inclination
angle $\alpha = \alpha (t)$ would be a possible way to explain the
phenomenon.

All the models described above are based on the simplifying assumption
that the kinetic energy $E_\textsc{k}$ of the body is only due to its
rigid rotation, i.e., $E_\textsc{k} = \frac12 I\Omega^2$. However, a
neutron star cannot be strictly considered as rigid and even though
the rotation is very slow from the point of view of relativistic
effects, with typical surface velocities of the order of\footnote{The
  Crab is among the fastest rotating isolated pulsars with known
  braking index, with velocity $v_\text{crab} \sim 1.2\times 10^{-2}
  c$.}  $v\simeq (10^{-4}- 10^{-2})c$, the shape should be allowed to
depend on time as this rotation may naturally induce a flattening of
the poles. In such a scenario, the kinetic energy acquires new
contributions that need to be taken into account.

It is the purpose of the present work to introduce a more complete
description taking into account the variation of the internal
potential energy of a self-gravitating body \cite{Lucas2015}. In
particular, we explore the consequences of allowing the star shape to
evolve in time, under its coupling with the rotation of the body. It
is assumed that energy can be lost in this process, a phenomenon that
could be relevant in the explanation of the measured slowing-down of
isolated pulsars.  Embedding our model in a general relativistic
context goes beyond the scope of this work, and as we consider the
quasi-rigid rotation of the star, we restrict attention to the
Newtonian weak field limit; given the slow velocities involved, we
expect this approximation to be meaningful.

In the next section, the basic assumptions of the model are described
and the coupled system of nonlinear differential equations governing
the evolution of the star are derived. The results of our numerical
analysis are presented in Sec.~\ref{estimates}, where some suggestive
solutions are studied. In particular, we show that a pulsar evolution
with $n < 3$ can easily be reproduced. A comparison between our
results and the available data describing the behavior of the Crab
pulsar is given in Sec.~\ref{crab}, before a few final and concluding
remarks in Sec.~\ref{final}.

\section{The model}
\label{model}

\begin{figure}[t]
    \includegraphics[scale=0.42]{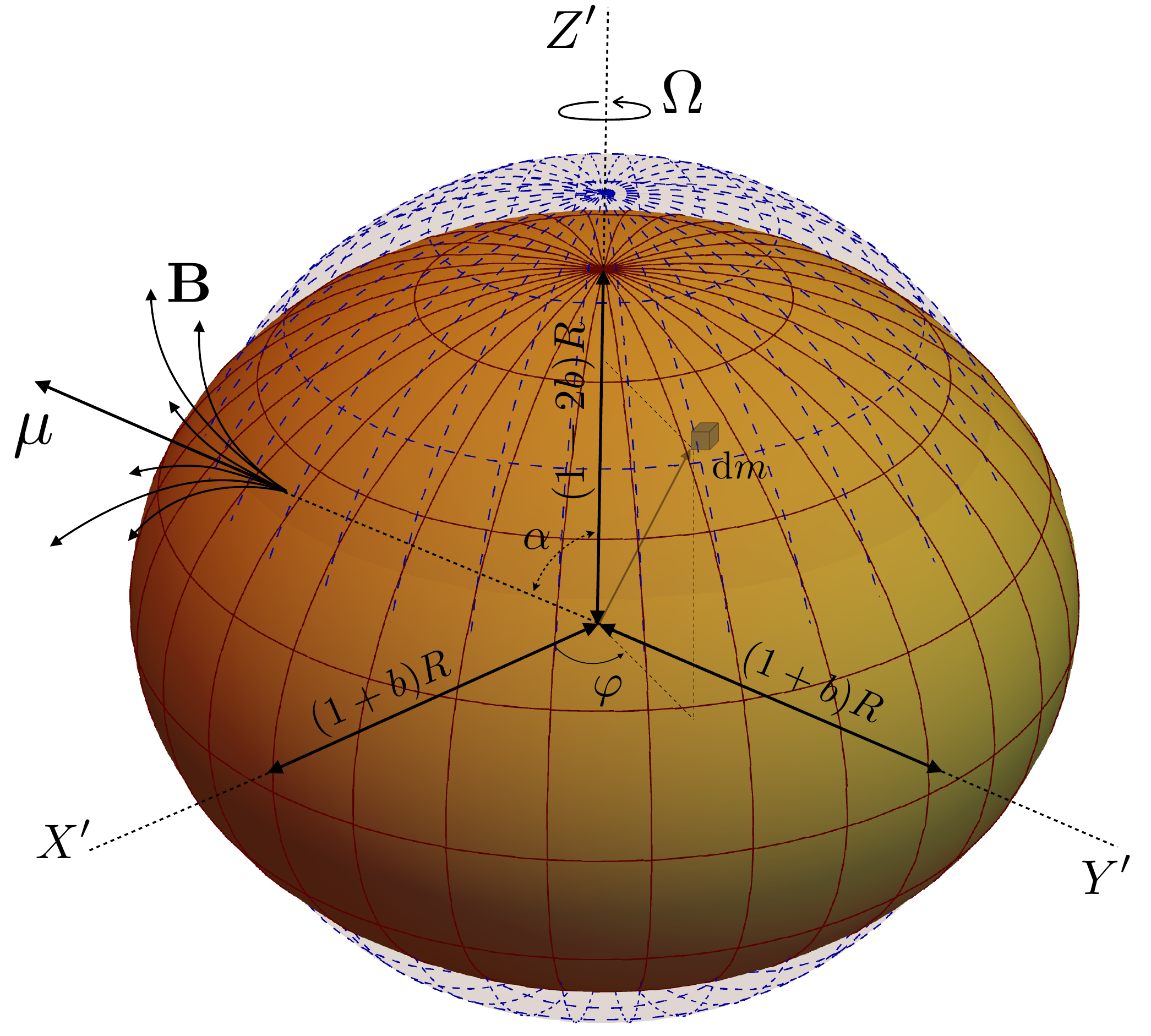}

    \caption{The pulsar geometric configuration: the neutron star is
      rotating with angular velocity $\Omega(t) = \dot\varphi$ around
      the $z$ axis while its magnetic axis is inclined by an angle
      $\alpha$ leading to precession of the field lines. The rotation
      induces a deviation from sphericity so the original sphere of
      radius $R$ (dashed grid) turns into an ellipsoid whose semi-axes
      are depend on time through the function $b(t)$ related to the
      ellipticity.}

    \label{ellipsoide}
\end{figure}

Suppose the star is described as a mass distribution that is slightly
deformed compared to a spherical body and is rotating around the
$z$-axis with time-dependent angular velocity $\Omega (t)$. We assume
the volume of the star to be that of the non rotating sphere $\frac43
\pi R^3$, its actual shape being ellipsoidal with two equal semi-axes
in the $(x,y)-$plane slightly larger than the sphere radius,
i.e. $(1+b)R$ (see Fig.~\ref{ellipsoide}).  Even for the large
velocities involved in a rotating neutron star, we do not expect large
deviations from sphericity and thus demand that $b\ll 1$. As a result,
the volume of the ellipsoid matches that of the sphere to second order
in $b$ provided the semi-axis in the direction of rotation is
$(1-2b)R$. An arbitrary element of mass $\dd m$ in the body will thus
be described by the set $\bm{R}' = \{X',Y',Z'\}$ such that
$$
\left(\frac{X'}{1+b}\right)^2 + \left( \frac{Y'}{1+b} \right)^2 +
\left(\frac{Z'}{1-2b}\right)^2 \le R^2.
$$
As discussed in the introduction, we expect the quantity $b$ may not
necessarily be constant, so we anticipate that $b=b(t)$. It is
convenient to use a coordinate system related to that of the embedding
sphere, i.e., the set $\{x,y,z\}$ such that $x^2+y^2+z^2 = r^2$, with
$r\le R$. That is, we implement the coordinate transformation $X'=
(1+b) x$, $Y'= (1+b) y$ and $Z'= (1-2b) z$. The volume element, as
expected, is $\dd V=\dd X' \dd Y'\dd Z'= \dd x\dd y\dd z +
\mathcal{O}(b^2)$.

Let us implement a rotation $\mathcal{R}_{\bm{\hat{z}}}(\varphi)$ to
the body about the $z$-axis by an angle $\varphi$. The position
$\bm{r}'$ of the mass element $\dd m$ is then given by the rotation
applied to its location $\bm{R}'$, i.e. $\bm{r}'(\varphi,b) =
\mathcal{R}_{\bm{\hat{z}}}(\varphi) \cdot \bm{R}'$. One therefore gets
\begin{equation}
\bm{r}' = \underbrace{\left( \begin{matrix}
\cos\varphi & \sin\varphi & 0\cr
-\sin\varphi & \cos\varphi & 0\cr
0 & 0 & 1
\end{matrix} \right)}_{\mathcal{R}_{\bm{\hat{z}}}(\varphi)}
\underbrace{\left( \begin{matrix}
1+b & 0 & 0\cr
0 & 1+b & 0\cr
0 & 0 & 1-2b
\end{matrix} \right)
\left( \begin{matrix}
x\cr
y\cr
z
\end{matrix} \right)}_{\bm{R}'},
\label{r}
\end{equation}
whose modulus we denote by $r'$.

Since we consider a pulsar, i.e. a rotating neutron star, one now
needs to assume from that point on that both the rotation angle
$\varphi$ and the flattening of the poles depend on time. The absolute
velocity, $v = \| \dd \bm{r}'/\dd t\|$, of each mass element in such
system is given by
\begin{equation}
v = \sqrt{(x^2+y^2+4z^2){\dot {b}}^2 + (x^2+y^2)\Omega^2 (1+b)^2},
\label{v}
\end{equation}
where $\Omega = \dot\varphi$ is the angular velocity of $\dd m$ around
the $z$-axis. The kinetic energy $E_\textsc{k}$ of the star is
obtained by integrating $\frac12 v^2\dd m$ over the whole body volume,
leading to
\begin{eqnarray}
E_\textsc{k} & = &\frac12 \dot{b}^2 \int_V (x^2+y^2+4z^2)\rho(r)\dd V
\nonumber\\ & & + \frac12 \Omega^2(1+b)^2 \int_V (x^2+y^2)\rho(r)\dd
V,
\label{Ek}
\end{eqnarray}
where $\rho(r)$ is the mass-density function that is supposed to
depend only on the distance to the origin.

The moment of inertia $I$ of the neutron star, seen as an idealized
spherical mass distribution rotating about the $z$-axis, is defined by
$$
I \doteq \int \left( x^2+y^2\right)\rho (r) \dd V,
$$
which, because of the spherical symmetry, is also expressible as
\begin{equation}
I = 2 \int \rho x^2 \dd V = 2\int\rho  y^2 dV =2\int\rho  z^2 \dd V.
\label{inertia}
\end{equation}
For the spherical approximation, it reads $I=\frac25 MR^2$, so that
Eq.~\eqref{beta} then implies $\beta\approx 5 B^2 R^4/(12 M)$.

The kinetic energy now reads
\begin{equation}
E_\textsc{k} = \frac{3}{2}I \dot b^2 + \frac{1}{2}I (1+2b)\Omega^2 +
\mathcal{O}(b^2).
\label{Ek2}
\end{equation}

As the system evolves, the body will be allowed to oscillate. Its
potential energy $E_\textsc{p}$ can be expanded about $b=0$, as
\begin{align}
E_\textsc{p} &\approx E_p(0) + \frac{1}{2}\kappa b^2 ,
\label{V}
\end{align}
where we have used that the potential energy is minimized for the
spherical configuration, so that $\left( \partial
E_\textsc{p}/\partial b \right)_{b=0} = 0$.  In Eq.~\eqref{V}, we
noted the elastic constant as
\begin{equation}
\kappa = \left( \frac{\partial^2 E_\textsc{p}}{\partial
  b^2}\right)_{b=0} = 3 I \gamma,
\label{gamma}
\end{equation}
thereby defining the coefficient $\gamma$. The leading contribution to
the elastic constant can be obtained by assuming the spherical
approximation, which leads to $\kappa \approx 24GM^2/(5R)$, such that
$\gamma \approx 4GM/R^3$.

Neglecting higher order terms, the Lagrangian of the system reads, up
to a constant,
\begin{align}
\mathcal{L} = \frac12 I\left[ 3\dot b^2 + (1+2b)\Omega^2
- 3\gamma b^2
  \right].
\label{L}
\end{align}
It should be noticed at this point that $\mathcal{O}(b^2)$ terms in
the kinetic contribution have been neglected because they are very
small when compared to $ \gamma I b^2$ coming from the potential
energy contribution. This corresponds to assume $\Omega^2 \ll GM/R^3$,
a condition that is related to the slow rotation Newtonian hypothesis,
satisfied for the physical system under consideration.

The Euler-Lagrange equations stemming from \eqref{L} must be
supplemented by dissipation terms~\cite{Levi1987,Lucas2015,Lucas2017},
in order to account for the radiation. They read
\begin{align}
3I\ddot b + 3I\gamma b - I\Omega^2 &= -\frac{\partial D}{\partial \dot
  b},
\label{eq1lag}
\\ \frac{\dd}{\dd t} \left[ I(1+2b)\Omega \right] &= -\frac{\partial
  D}{\partial \dot \varphi},  
\label{eq2lag}
\end{align}
where the dissipation function $D$ is here related to the dipole
radiation emission and the damping of the body oscillations.  Our
simplified model relies on internal dissipation processes associated
with the quadrupole moment tensor, and we demand that the oscillations
have a small amplitude such that they should remain linear in their
time derivative.  These requirements can be achieved with the
following prescription for the dissipation function
\begin{equation}
D = \frac{1}{6}\mu^2\Omega^4 \sin^2\!\alpha + \frac{3}{2}\sigma I\,
\dot b ^2,
\label{dforce}
\end{equation}
thereby defining our final phenomenological parameter $\sigma$.

Now, defining the total energy $E=E_\textsc{k}+E_\textsc{p}$, and
using the above results, it is straightforward to evaluate the energy
losses, namely
\begin{align}
\dot E = - \dot\varphi \frac{\partial D}{\partial\dot\varphi} -\dot b
\frac{\partial D}{\partial \dot b} = -\beta I \Omega^4 - 3\sigma I
\dot b^2,
\label{energyloss}
\end{align}
which is the equation that governs the energy balance of the system.

In a scenario where MDR is the only process behind the loss of energy
of a pulsar, Eq.~(\ref{eq1}) would hold and the external torque
$\tau_\text{ext} = -\beta I \Omega^3$ would be the only responsible
for the star slowdown.  However, in the more complete scenario under
investigation in the present work, the evolution of the system is
governed by the set of coupled equations of motion given by
Eqs.~\eqref{eq1lag} and \eqref{eq2lag} which, after inserting
Eq.~\eqref{dforce}, can be presented in the more compact form as
\begin{subequations}
\begin{align}
&\ddot b + \sigma \dot b + \gamma b = \frac{1}{3}\Omega^2,
\label{motion-eq1}
\\
&\dot\Omega = -\frac{2\Omega \dot b}{(1+2b)} -\frac{\beta}{(1+2b)}\Omega^3,
\label{motion-eq2}
\end{align}
\label{motion}%
\end{subequations}
where the three parameters $\beta$, $\gamma$ and $\sigma$ in the above
equations are given by Eqs.~\eqref{beta}, \eqref{gamma} and
\eqref{dforce}. From the point of view of physical units, they are
expressed respectively in $\mbox{s}$ ($[\beta]=T$), $\mbox{s}^{-2}$
($[\gamma]=T^{-2}$) and $\mbox{s}^{-1}$ ($[\sigma]=T^{-1}$).

Before closing this section, a few words about angular momentum
conservation are in order. First, in our model, the quantity
$\mathcal{I} \doteq I(1+b)^2$ is identified as the time-dependent
effective moment of inertia of the body, thus making
Eq.~\eqref{motion-eq2} the equation of motion relating the total
angular momentum $L={\cal I}\Omega$ with the external torque produced
by the emission of MDR. Naturally, in the absence of external torque,
the angular momentum is a conserved quantity, i.e., when
$\tau_\text{ext}=-\partial D/\partial \dot \varphi=0$. As expected,
internal processes, as those described by the second term in the rhs
of Eq.~\eqref{dforce}, do not interfere with the angular momentum
conservation law. Finally, it should be emphasized that when the
moment of inertia is allowed to vary with time, $\frac12\mathcal{I}
\Omega^2$ will not be the only contribution to the kinetic energy of
the body, as clearly emphasized by Eq.~\eqref{Ek2}. The time evolution
of the angular momentum is governed by $\Omega(t)$, and also by
$\mathcal{I}(t)$ through $b(t)$. These functions are solutions of the
coupled differential equations of motion \eqref{motion} that naturally
follow from the Lagrangian method.

\section{Modelling a pulsar slowdown}
\label{estimates}

Quantities like the mass of the pulsar, its radius, or the strength of
the field at its magnetic pole are not known with great precision, and
these values can also be model dependent. For instance, the mass of
the Crab pulsar is usually taken to be approximately $M_\text{crab}
\simeq 1.4 \,M_\odot$, with $M_\odot$ the solar mass. The goal of this
work is to test if our theoretical model is able to produce acceptable
solutions to the problem of pulsars slowdown, i.e., if a braking index
less than 3 is possible when the oscillations described by $b(t)$ are
taken into account.

Using the results obtained in the last section, the parameters $\beta$
and $\gamma$ can be conveniently expressed in terms of $M_\odot$ and
the typical values for the radius and the magnetic dipole field of a
certain class of known pulsars, namely
\begin{align}
    \beta & \approx 6.190 \times 10^{-19} \sin^2\alpha
    \left(\frac{M_\odot}{M}\right) \left(\frac{B}{10^8{\rm
        T}}\right)^2\left(\frac{R}{10{\rm km}}\right)^4 {\rm s}
    \nonumber \\ \gamma & \approx 5.307 \times 10^8
    \left(\frac{M}{M_\odot}\right)\left(\frac{10{\rm km}}{R}\right)^3
         {\rm s}^{-2}, \nonumber
\end{align}
in which we used the expressions for $I$ and $\mu$ assuming a
spherical star.

In the subsequent numerical calculations, we assume specific values
for the neutron star model: we fix the radius $R= 1.674 \times 10^4
\,{\rm m}$, consider that the magnetic dipole generating a field
amplitude $B= 1.428\times 10^{9} \,{\rm T}$ (this denotes the
magnitude of the field at the pole of the star \cite{shapiro}), and
allow for a misalignment with the rotation axis by the angle $\alpha =
(\pi/4)\, {\rm rad}$. Although these values are chosen here merely for
convenience, they happen to describe with reasonable accuracy several
known isolated pulsars.

\begin{figure}[!t]
    \includegraphics[scale=0.41]{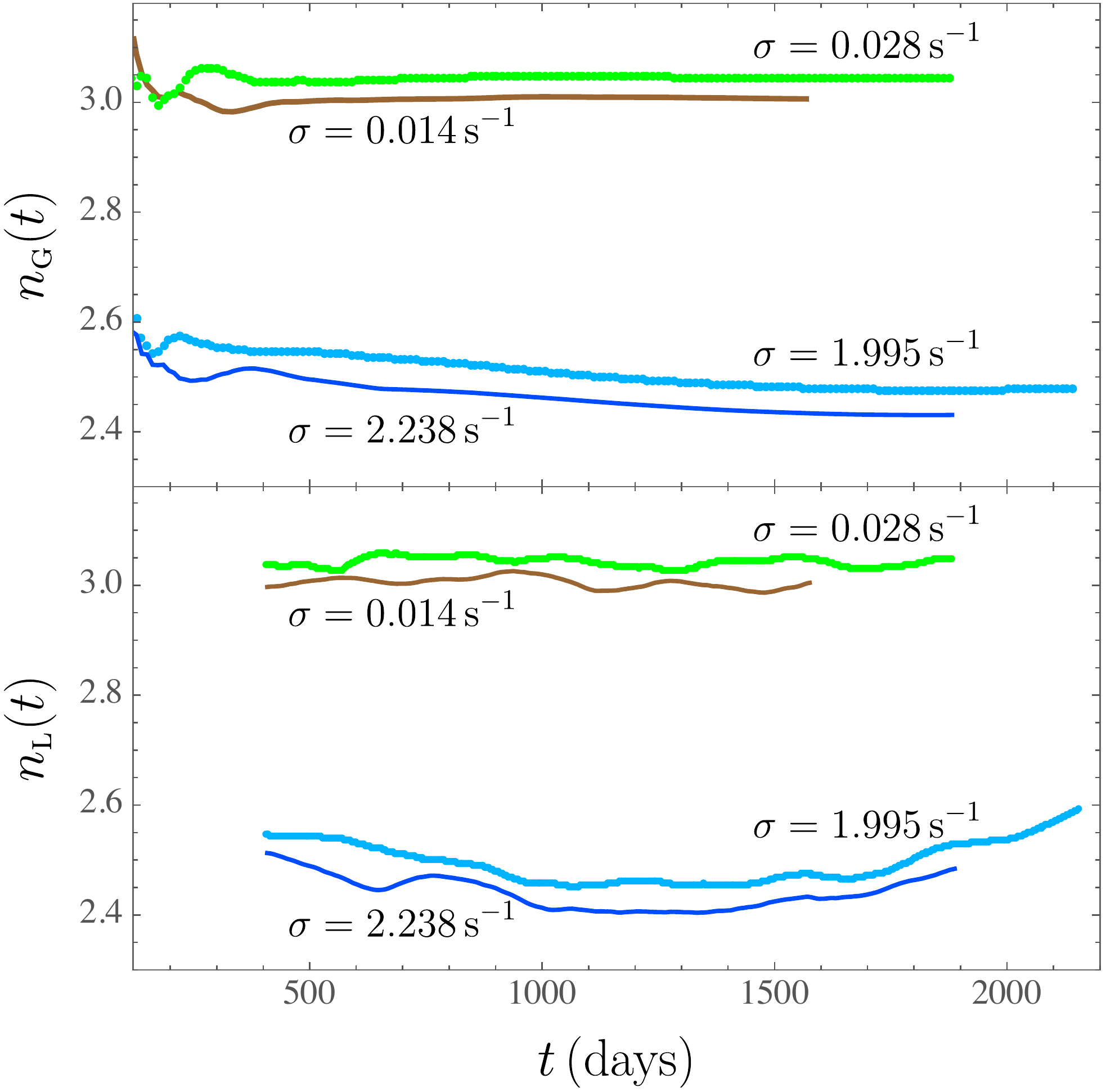}

    \caption{Global [upper panel, Eq.~\eqref{globalbrak}] and local
      [lower panel, Eq.~\eqref{localbrak}] braking indices calculated
      from the simulation data for some representative values of the
      dissipation parameter $\sigma$. Note that there is no direct
      relationship between the magnitude of the dissipation process
      and the order of the braking indices when small variations of
      $\sigma$ are considered. However, on average, a more intense
      dissipation process (larger $\sigma$ values) leads to smaller
      values for the braking index.}

    \label{fig-n-positive}
\end{figure}

\begin{figure}[!t]
    \includegraphics[scale=0.4]{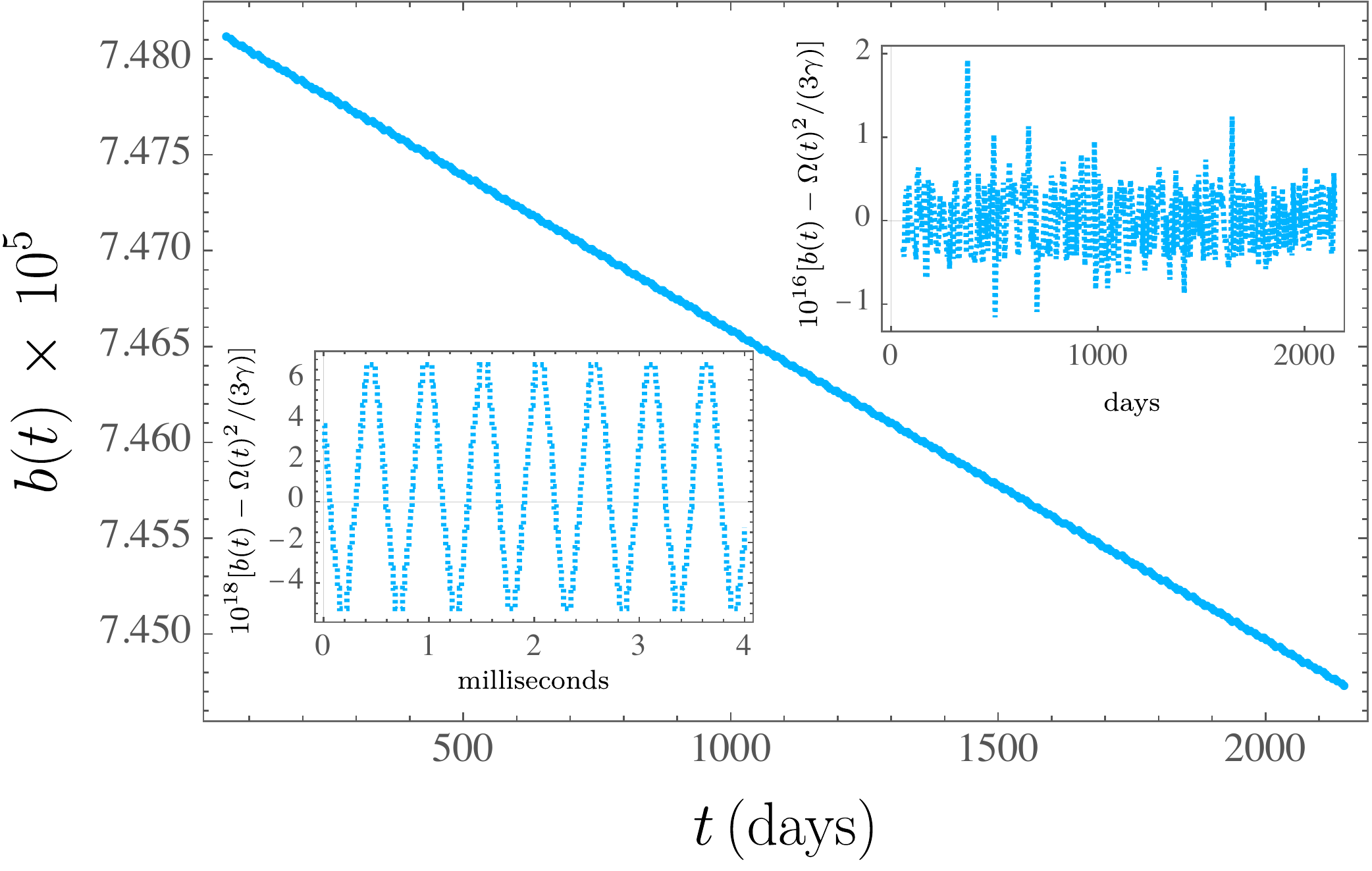}

    \caption{Time evolution of $b(t)$ as the pulsar slows down. The
      upper inset depicts the behavior of the normalized $b(t)$,
      obtained when its time-evolving equilibrium point
      $\Omega(t)^2/(3\gamma)$ is subtracted, thus showing that it is a
      highly oscillating function in time but on a small amplitude
      scale. The lower inset presents a close-up of the normalized
      curve (about day 2000) in a scale of milliseconds, showing that
      the oscillations occur with a well defined frequency,
      approximately the natural frequency $\sqrt{\gamma}/(2\pi)$ of
      the harmonic oscillator described by Eq.~\eqref{eq1lag}. }

    \label{fig-b}
\end{figure}

Following the model described in the previous section, the evolution
of the system is governed by the coupled nonlinear differential
equations given by Eqs.~\eqref{motion}. The initial angular velocity
is set to be $\Omega(0) = 188.5\, {\rm rad}\cdot {\rm s}^{-1}$.  The
initial deformation of the body, $b(0)$, is assumed to be the
equilibrium value of $b(t)$ in Eq.~(\ref{motion-eq2}), i.e., $b(0) =
\Omega(0)^2/(3\gamma) \approx 7.482\times 10^{-5}$, for which $\dot
b(0)$ was set to zero.

With the above assumptions, it follows that the underlying parameters
are given by $\beta \approx 3.542\times 10^{-16}\, {\rm s}$ and
$\gamma \approx 1.583\times 10^{8}\, {\rm s}^{-2}$. The remaining
parameter $\sigma$ is associated to the dissipation processes during
the oscillations of the quadrupole moment of the body, and will be
adjusted in the simulations in order to obtain the braking index of
the pulsar.

In a fashion similar to that present in the analysis in the existing
literature for the calculation of the braking index of the Crab
pulsar~\cite{lyne2015}, we consider here the following method, that
can be applied to both simulated or measured data:
\begin{enumerate}[$(i)$]
    \item Let $(t, \Omega) = (t_k, \Omega_k)$, for $k = 1, \cdots, N$,
      denote the complete time series for the angular velocity of the
      pulsar;
    \item For each $j$ such that $1 < j \leq N$, the {\em global
      braking index} at time $t_j$ is computed by fitting the points
      $(t_k, \Omega_k)$, where $k = 1, \cdots, j$, with the 3rd-degree
      polynomial
     $$P_{\textsc{g};j}(t) = \sum_{k=0}^3 a_{k;j} \left( t -
      t_{\textsc{g};j} \right)^k,$$ where $t_{\textsc{g};j} = \frac12
      \left( t_j-t_0\right)$ is the half time of the interval
      $[0,t_j]$, and the corresponding braking index at $t_j$ is given
      by Eq.~\eqref{brake}, reading here
    \begin{equation}
       n_{\textsc{g}}(t_j) =
       \frac{\ddot{P}_{\textsc{g};j}(t_{\textsc{g};j})
         P_{\textsc{g};j}(t_{\textsc{g};j})}{\dot{P}_{\textsc{g};j}^2(t_{\textsc{g};j})}
       = 2\frac{a_{2;j} a_{0;j}}{a_{1;j}^2};
       \label{globalbrak}
    \end{equation}
    \item The {\em local braking index} at time $t_j$ must be
      determined over each data subset with a fixed size $\Delta N \in
      \mathbb{N}$ as illustrated below:
      
\begin{tikzpicture}[scale=3.8]
    \draw (-0.1,0.) node {$\bullet$} ;
    \draw (-0.1,0.) node[below left]{$1$}
      -- (1.7,0.) node[below right]{$N$} ;
    \draw (0.6,0.) node {$|$} ;
    \draw (0.6,0.) node [above] {$\null^{(j-\Delta N)}$} ;
    \draw (0.85,0.05) node [above] {$t_{\textsc{l};j}$} ;
    \draw (0.85,0.) node {$|$} ;
    \draw (1.1,0.) node {$|$} ;
    \draw (1.1,0.) node [above] {$\null^j$} ;
    \draw (1.7,0.) node {$\bullet$} ;
    \draw (0.85,-0.03) node [below] {$\underbrace{\hskip 1.8cm}_{\Delta N}$} ;
\end{tikzpicture}

      In this case, for each $\Delta N < j \leq N$, the local braking
      index at time $t_j > t_{\Delta N}$ is computed by fitting the
      points $(t_k, \Omega_k)$, where $k = j - \Delta N, \cdots, j$,
      with the 3rd-degree polynomial
     $$P_{\textsc{l};j}(t) = \sum_{k=0}^3 a_{k;j} \left( t - t_{\textsc{l};j} \right)^k,$$
     where
      $t_{\textsc{l};j} = \frac12 \left( t_{j-\Delta N} + t_j\right)$ is the
     center of the interval $[t_{j -\Delta N},t_j]$, and the corresponding local
      braking index is again given by Eq.~\eqref{brake}, namely
    \begin{equation}
       n_{\textsc{l}}(t_j) =
       \frac{\ddot{P}_{\textsc{l};j}(t_{\textsc{l};j})
         P_{\textsc{l};j}(t_{\textsc{l};j})}{\dot{P}_{\textsc{l};j}^2(t_{\textsc{l};j})}
       = 2\frac{b_{2;j} b_{0;j}}{b_{1;j}^2}.
       \label{localbrak}
    \end{equation}
\end{enumerate}

An extrapolation-algorithm, based on the explicit midpoint rule, with
stepsize control and order selection (see Section II.9 from
Ref.~\cite{Hairer1993}) was used to numerically integrate the coupled
system described by Eqs.~(\ref{motion-eq1}) and (\ref{motion-eq2}),
leading to the results depicted in Figs.~\ref{fig-n-positive} to
\ref{fig-b}. The integration spans a time window of about 5 years,
which was enough to obtain solutions with stable braking indices. In
fact, after a short time of instability, $n(t)$ eventually behaves as
a slowly evolving function of time, as it can be confirmed by direct
inspection of Fig.~\ref{fig-n-positive}, where some solutions
presenting positive braking indices were selected.

\begin{figure}[!t]
    \includegraphics[scale=0.4]{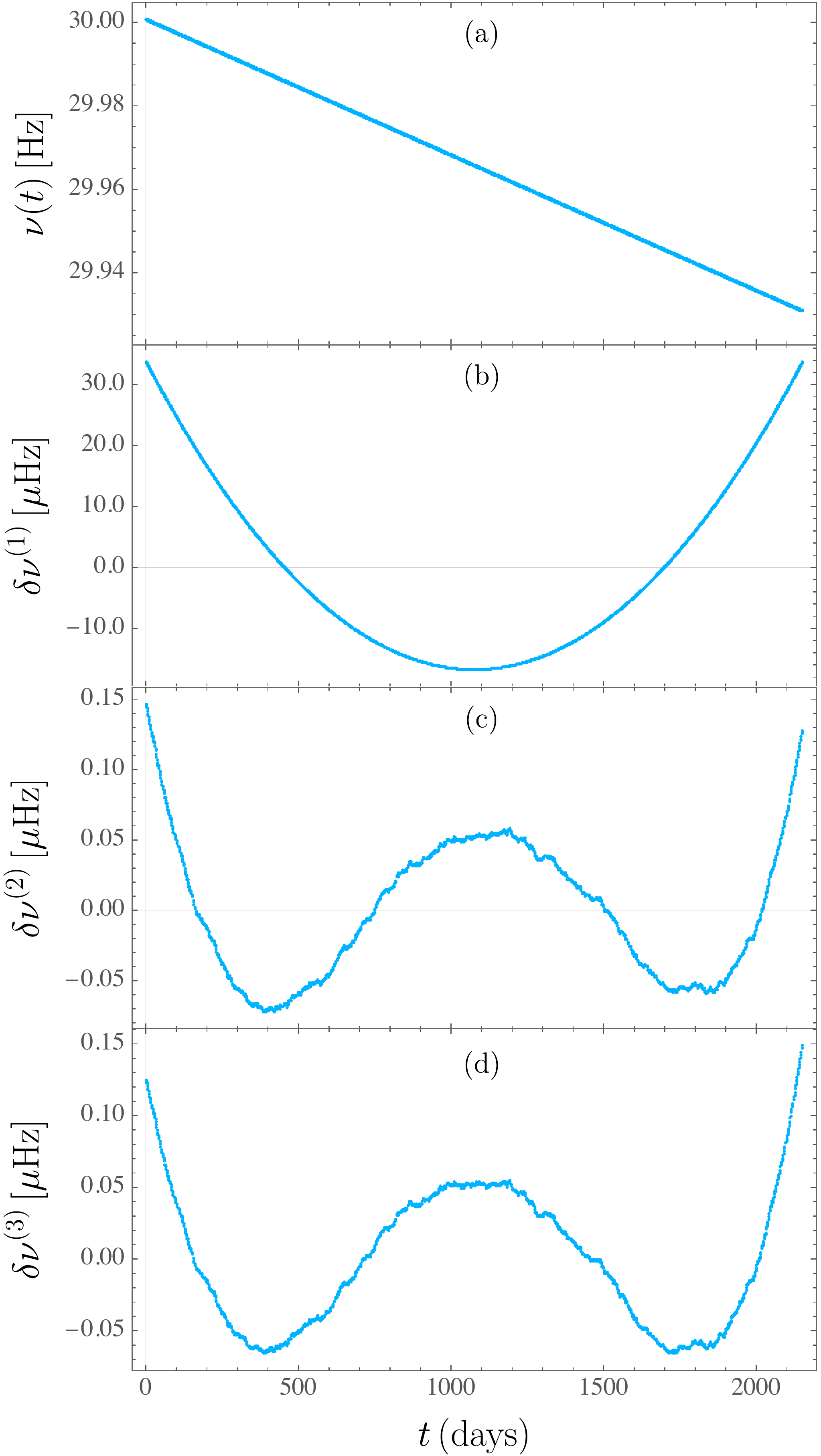}

    \caption{Rotational frequency (a) and residuals of the first (b),
      second (c) and third-order (d) shown as functions of time (in
      days) for the simulation data corresponding to a dissipation
      parameter $\sigma=1.995\, \mbox{s}^{-1}$. The global braking
      index corresponding to this simulation data, after about 6 years
      of integration time, achieves a value of approximately 2.5, as
      can be inferred by direct inspection of
      Fig.~\ref{fig-n-positive}. The initial frequency in this
      simulation was $30.00 \mbox{Hz}$.}

    \label{fig-Omega}
\end{figure}

The magnitude of the dissipation process associated with the
quadrupole oscillations is dominant in determining the behavior of the
braking index of the system. Processes for which $\sigma$ is of the
order of $0.01\,\mbox{Hz}$ lead to braking indices around $n=3$, which
is the expected result when the pulsar's rotational energy is taken
away only by means of magnetic dipole radiation. However, for higher
values of $\sigma$, richer scenarios appear, as shown in
Fig.~\ref{fig-n-positive}. In particular, when $\sigma\approx 2\,
\mbox{Hz}$, the solutions exhibit braking indices around 2.5.  Small
variations of $\sigma$ lead to different solutions for $n(t)$. On the
other hand, this function does not seem to be very sensitive to small
variations of the other parameters. The local behavior of the braking
index, depicted in the lower panel of Fig.~\ref{fig-n-positive}, was
obtained using a moving average ($\Delta N$) of 400 days, which
explains why it starts after the global index (upper panel).  When a
sufficiently high dissipation process is taken into account, the
simulations suggest that even negative braking indices are possible
solutions. This is an aspect that deserves further examination.

The behavior of the angular velocity is very similar for all solutions
examined in Fig.~\ref{fig-n-positive}. If the curves corresponding to
the angular velocities for these models were included in a same plot,
almost no visual difference would be seen. In fact, it can be shown
that for any instant of time in the simulations, the difference
between the angular velocities of any of these solutions is smaller
than $10^{-5}\mbox{Hz}$.

The behavior of $b(t)$ for the model with $\sigma =1.995\,
\mbox{s}^{-1}$ is shown in Fig.~\ref{fig-b}. In the plot scale it
looks like a slowly decreasing monotonic function of time. However, a
more detailed examination shows that $b(t)$ is a highly oscillatory
function around the time-dependent equilibrium point
$\Omega(t)^2/(3\gamma)$, as highlighted in the inserts. Indeed, due to
their mutual coupling, both $b(t)$ and $\Omega(t)$ decompose into a
slow monotonically decreasing component and a fast (and tiny)
oscillatory component; the slow component can be extracted out by
calculating the difference $b(t) - \Omega(t)^2/(3\gamma)$.  Thus, as
the system loses energy by means of MDR emission and oscillation
damping, it will slow-down its rotation frequency and also the
amplitude of the oscillations.

The evolution of the rotation frequency corresponding to the model
with $\sigma = 1.995 \,\mbox{s}^{-1}$ is depicted in
Fig.~\ref{fig-Omega}(a), which is a solution presenting a braking
index of approximately $2.5$. The other panels in the figure show the
residuals of the first, second and third order, which were obtained
following the standard procedure (see for instance the analysis for
the Crab pulsar \cite{lyne2015}): the data set is fitted by means of a
$k$-degree polynomial, which can be written as $\nu(t) =
\sum_{i=0}^{k} c_i (t-t_0)^i + \delta \nu^{(k)}$, where $t_0$ is
chosen, for instance, to be the medium time of the data set, the
coefficients $c_i$ are obtained by the fitting procedure, and the
time-dependent function $\delta \nu^{(k)}$ is the $k$-th order
residual obtained when the $k$-th order fitting polynomial is
subtracted from the data. For instance, Fig.~\ref{fig-Omega}(b)
depicts the first-order residual $\delta\nu^{(1)} =
[\nu(t)]_{\mbox{\tiny data}} - [c_0 + c_1 (t-t_0)]$, where $t_0 =
9.299\times 10^7\, \mbox{s} \approx 1076 \,\mbox{days}$, $c_0 \approx
29.97\, \mbox{Hz}$, and $c_1 \approx - 3.757 \times 10^{-10}
\,\mbox{s}^{-2}$.

The residuals at second and third order exhibit some irregularities,
which cannot be attributed to numerical errors. 
They can be interpreted as very short moments in
time during which the angular momentum is suddenly changed 
before the star returns to its original state.
That could be interpreted as micro-glitches: a realistic model
would describe for instance various rotating shells, all of which
would be subject to equations similar to those presented here and
somehow interacting. Could such a more elaborate model enhance this
phenomenon to the level of the observed glitches?

\section{A note about glitches: the behavior of the Crab pulsar}
\label{crab}

Having described our simple model, one wants to compare with the
existing data relevant to the dynamical range under investigation. The
best example one can think of is provided by the enormous amount of
data available concerning the Crab pulsar.

\begin{figure}[!h]
    \includegraphics[scale=0.4]{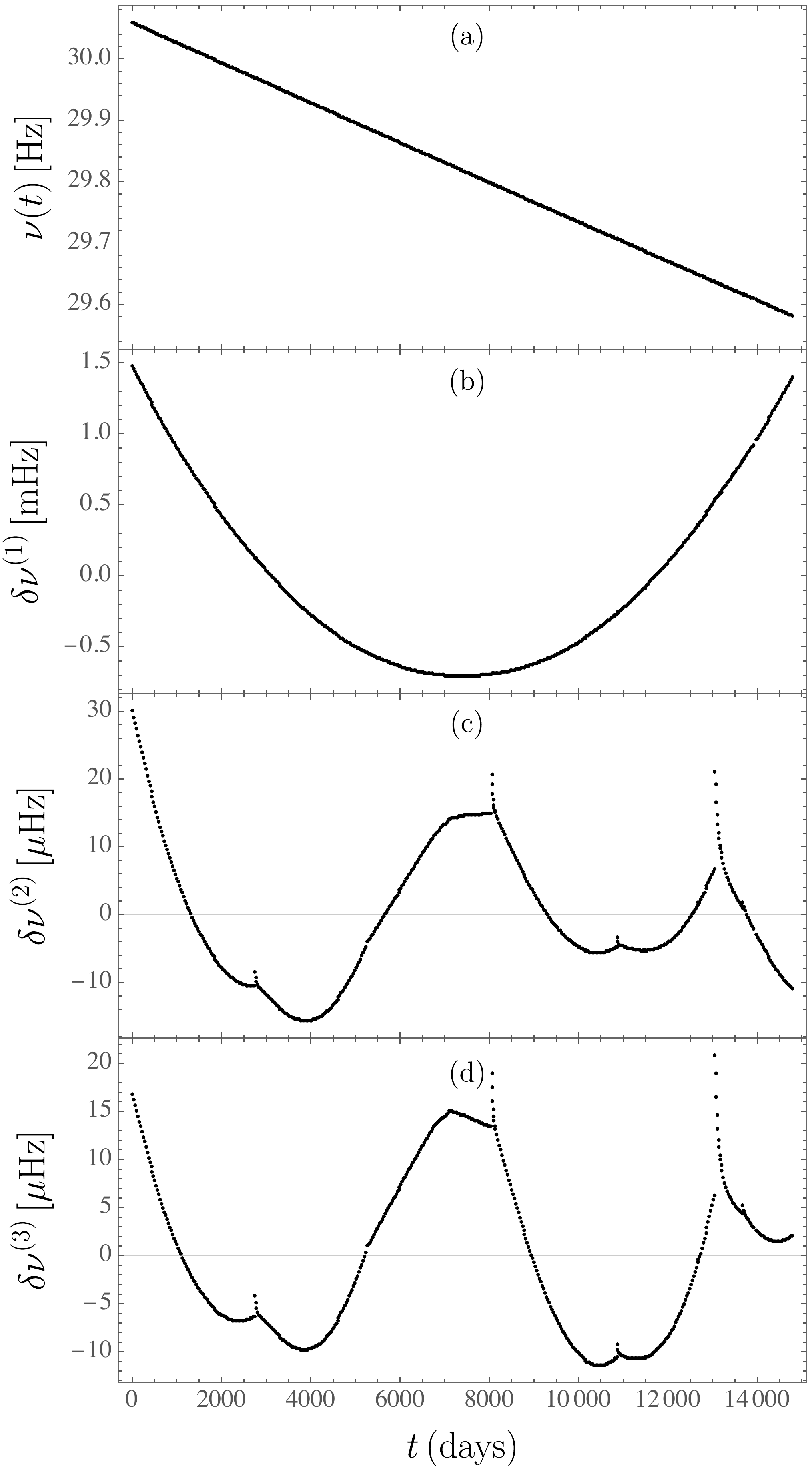}
    \caption{Rotation frequency and residuals of the Crab pulsar,
      according to data collected from 1982 onwards
      \cite{lyne1993,lyne2015}. The starting time in this figure
      corresponds the MJD 45015, for which the measured frequency was
      $\nu = 30.0592241133\, \mbox{Hz}$. The residuals of second and
      third orders, depicted in panels (c) and (d), respectively,
      clearly show a rich glitch activity of the pulsar in this
      period.}
    \label{residuals-Crab}
\end{figure}

The Crab pulsar (PSR B0531+21) is an isolated neutron star whose
angular velocity deceleration has been measured since the
1970s. Monthly spaced pulsar timing measurements have been taken by
Jodrell Bank Observatory since 1982 \cite{lyne1993}.  In
Fig.~\ref{residuals-Crab}(a), the rotation frequency measured for the
Crab pulsar is shown as a function of time, from MJD 45015 (February
15, 1982) to MJD 59806 (August 15, 2022)~\cite{lyne1993}. The
residuals are shown from top to bottom [panels (b) to (d)], where the
coefficients of the third-order polynomial fitting are $t_0 = 6.389712
\times 10^{8}\, \mbox{s} = 7395.5 \,\mbox{days}$, $c_0 \approx 31.36\,
\mbox{Hz}$, $c_1 \approx - 4.205 \times 10^{-10} \,\mbox{s}^{-2}$,
$c_2 \approx 6.728 \times 10^{-21} \,\mbox{s}^{-3}$, and $c_3 \approx
- 1.260 \times 10^{-31} \,\mbox{s}^{-4}$.

This system has occasional glitches, in which the star is spinned-up
for a short period of time and returns to its former rotation
frequency within an interval of about 20 days. The glitch activities
can be clearly seen in Fig.~\ref{residuals-Crab}(c) and
\ref{residuals-Crab}(d). They contribute massively to the global
braking index. The global and local indices agree with the constant
value $n\approx 2.5$ if and only if they are computed in an interval
not containing a glitch. However, the global braking index decreases
monotonically in time as multiple intervals are included in the data
set so that its final value reaches $n \approx 2.3$ \cite{lyne2015}.

\begin{figure}[!t]
    \includegraphics[scale=0.41]{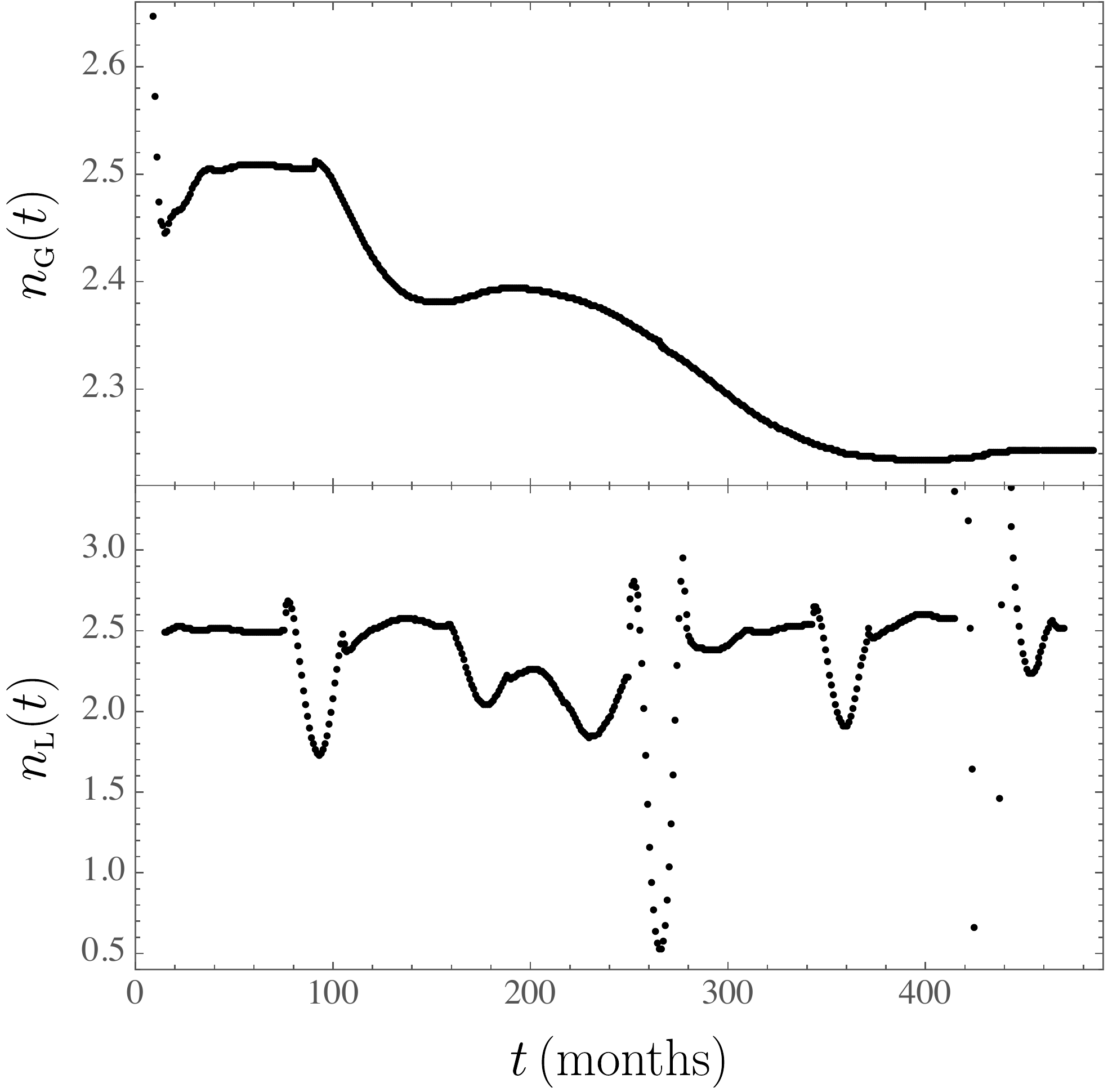}

    \caption{Braking index as function of time for the Crab
      pulsar. Here the index is calculated using the global and local
      procedures discussed in the last section, as depicted in upper and lower panels, 
      respectively. A moving average with $\Delta N = 30$
      was here used to obtain the local index.}

    \label{fig-Crab-n}
\end{figure}

Global and local behaviors of the braking index with time are shown in
Fig.~\ref{fig-Crab-n}, upper and lower panels respectively.  In
particular, the local index as a function of time (in months) is shown
in lower panel of Fig.~\ref{fig-Crab-n}, where the moving average was calculated
using a window containing 30 successive measurements.  It can be seen
that after each glitch, the braking index returns approximately to the
value it had before the glitch, and this happens in less than a
month. However, its influence in the local braking index calculation
goes way longer, an aspect that is dependent on the choice of $\Delta
N$.

\section{Final remarks}
\label{final}

In this work, we explored the idea that as MDR is emitted by an
isolated pulsar, its energy is continuously driven away, causing a
slow-down of its spin, and a possible modification of the shape of the
star. Although the MDR emission is largely accepted in the literature,
adding a perturbation in its ellipsoidal shape by means of small
oscillations has never been considered. As the star cannot be strictly
rigid, oscillations are expected: they are produced almost in a
stationary regime as it is linked to the spin slow-down process. These
oscillations must be dissipated by internal phenomena leading to a
secondary form of energy loss by the star. The possible mechanisms
behind such dissipation of energy were not considered in details in
this work. Instead, it was assumed that the effect is described by an
effective damping process that is dependent of the velocity square of
the quadrupole moment oscillations, leanding to a forced (by means of
MDR emission) and damped linear differential equation governing the
evolution of the body oscillations. In planetary tide theory
terminology, this equation describes a Kelvin-Voigt damping of the
quadrupole moment oscillations, endowed with a deformation inertia
term \cite{Lucas2018}. Additionally, the equation of motion for the
angular momentum of the star generalizes previous treatments where the
contribution due to the time-dependent moment of inertia was
ignored. As a consequence of this description, solutions presenting
braking indices below the predicted value for a pure MDR model
($n_\textsc{mdr}=3$) were found by means of numerical calculations. In
particular, we found that there exist choices for the phenomenological
parameters for which the solutions exhibit values similar to those
measured in isolated pulsars.

It should be noticed that the braking index calculated by means of the
global method is highly dependent on the initial conditions of the
system. Furthermore, if glitches occur during the evolution, as is the
case in most of the isolated pulsars, they can significantly
contribute to the value of this index. This aspect can be appreciated,
for instance, in the case of the Crab pulsar \cite{lyne2015}, where
the value of $n$ calculated by means of the local method results in
$n_\textsc{l} = 2.51$, while the global method leads to $n_\textsc{g}
= 2.34$. If the data set is restricted to the period from 1982
onwards, the global method would result in a different value, while
the local index would not be significantly affected, as discussed in
the previous section. This suggests that the local method provides a
more robust index to describe the slow-down of isolated pulsars.

The exact reason behind the occurrence of glitches in an isolated
pulsar is still a matter of investigation. Most likely, they are
associated with redistribution of mass in short time intervals
activated by resonance phenomena throughout the evolution of the
system. After a glitch, the system approximately returns smoothly to
its former state.  In the idealized model we investigated here, it is
assumed that the shape of the star evolves in time, as governed by the
oscillating function $b(t)$. Thus, after the initial transient, the
ellipsoid describing the star's surface will oscillate with an almost
constant amplitude. In this scenario, localized sub-micro glitches are
expected to occur all the time, as suggested by the zigzags in the
second and third-order residuals appearing in Fig.~\ref{fig-Omega}. A
more elaborate model could shed more light on this important
issue. Consider a multi-layer model for instance. In such a model,
resonance effects between the different layer oscillations could lead
to significant mass redistribution, and possibly to macroscopic
glitches, that would then have to be compare to those observed in
isolated pulsars. This is an issue that deserves investigation.

Among the possible applications of this work, it should be mentioned
that the experimental knowledge of the rotation frequency curve of a
given pulsar could be used as a starting point to find the best set of
physical parameters behind its behavior. It should be noted however
that the damping effects over the oscillations due to emission of
thermal radiation and quadrupole gravitational radiation for instance,
are not yet fully understood for such systems and also deserve further
investigation. Models assuming different forms for the dissipation
function and its implications in the possible values of $n$ could be
of great value in such investigations.

\vspace{0.8in}
\begin{acknowledgments}
V. A. D. L. is supported in part by the Brazilian research agency CNPq
(Conselho Nacional de Desenvolvimento Cient\'{\i}fico e Tecnol\'ogico)
under Grant No. 305272/2019-5. L. S. R. is supported in part by CFisUC
projects (UIDB/04564/2020 and UIDP/04564/2020), and ENGAGE SKA
(POCI-01- 0145-FEDER-022217), funded by COMPETE 2020 and FCT,
Portugal, and also FAPEMIG (Funda\c{c}\~ao de Amparo \`a Pesquisa no
Estado de Minas Gerais) under Grant No. RED-00133-21.
\end{acknowledgments}

\end{document}